%% file: main.tex
\documentclass[sigplan,screen]{acmart}
\usepackage{hyperref}  
\usepackage{booktabs} 
\usepackage{balance}
\usepackage{subcaption}
\usepackage{multirow}
\usepackage{graphicx}
\usepackage{float}

\setcopyright{none}
\acmDOI{}
\acmYear{2018}
\copyrightyear{2018}
\acmISBN{--}
\acmConference{Preprint}{March}{2020}

\renewcommand\footnotetextcopyrightpermission[1]{} 

\begin{document}
\title{Context-Aware Parse Trees}

\author{
 Fangke Ye$^{* 1,2}$ \hspace{5mm}
 Shengtian Zhou$^{* 1}$ \hspace{5mm} 
 Anand Venkat$^{1}$ \hspace{5mm}
 Ryan Marcus$^{1,3}$ \hspace{5mm} \\
 Paul Petersen$^{1}$ \hspace{5mm}
 Jesmin Jahan Tithi$^{1}$ \hspace{5mm} 
 Tim Mattson$^{1}$ \hspace{5mm} 
 Tim Kraska$^{3}$ \hspace{5mm} \\
 Pradeep Dubey$^{1}$ \hspace{5mm}
 Vivek Sarkar$^{2}$ \hspace{5mm}
 Justin Gottschlich$^{1,4}$
}
\affiliation{\vspace{2mm} $^1$Intel Labs \quad $^2$Georgia Tech \quad $^3$MIT CSAIL \quad $^4$University of Pennsylvania}

\renewcommand{\shortauthors}{Ye and Zhou et al.}

\input{abstract}

\keywords{code similarity, program synthesis, machine programming, software development, software maintenance}

\maketitle

\input{introduction}

\input{capt}

\input{experiments}

\input{related}

\input{conclusion}

\balance
\bibliographystyle{ACM-Reference-Format}
\bibliography{ml}

\end{document}

%% file: abstract.tex
\begin{abstract}

The simplified parse tree (SPT) presented in Aroma, a state-of-the-art code recommendation system, is a tree-structured representation used to infer code semantics by capturing program \emph{structure} rather than program \emph{syntax}. This is a departure from the classical abstract syntax tree, which is principally driven by programming language syntax. While we believe a semantics-driven representation is desirable, the specifics of an SPT's construction can impact its performance. We analyze these nuances and present a new tree structure, heavily influenced by Aroma's SPT, called a \emph{context-aware parse tree} (CAPT). CAPT enhances SPT by providing a richer level of semantic representation. Specifically, CAPT provides additional binding support for language-specific techniques for adding semantically-salient features, and language-agnostic techniques for removing syntactically-present but semantically-irrelevant features. Our research quantitatively demonstrates the value of our proposed semantically-salient features, enabling a specific CAPT configuration to be 39\% more accurate than SPT across the 48,610 programs we analyzed.

\vspace{3mm}
\noindent
{\emph{* Lead Authors.}}

\end{abstract}


%% file: introduction.tex
\vspace{-3mm}
\section{Introduction}

\emph{Machine programming} (MP), as defined by Gottschlich et al. in \emph{"The Three Pillars of Machine Programming,"} is any system that automates some aspect of software development~\cite{gottschlich:2018:mapl}. An open research challenge in MP is how to build effective automated code similarity systems. The potential use cases for such code similarity systems ranges from code recommendation to automated software bug patching, to name a few~\cite{dinella:2020:hoppity, luan:2019:oopsla, allamanis:2018:iclr, pradel:2018:oopsla, bhatia:2018:icse, bader:2019:oopsla, barman:2016:oopsla}. Yet, as others have noted, the correct \emph{structural representation} for such a code similarity system remains unclear~\cite{luan:2019:oopsla, odena:2020:iclr, ben-nun:2018:neurips, alon:2018:pldi, alon:2019:popl, tufano:2018:msr, alon:2019:iclr, zhang:2019:icse, allamanis:2018:acm}.

\begin{figure}[htpb]
\begin{center}
\includegraphics[width=0.8\columnwidth]{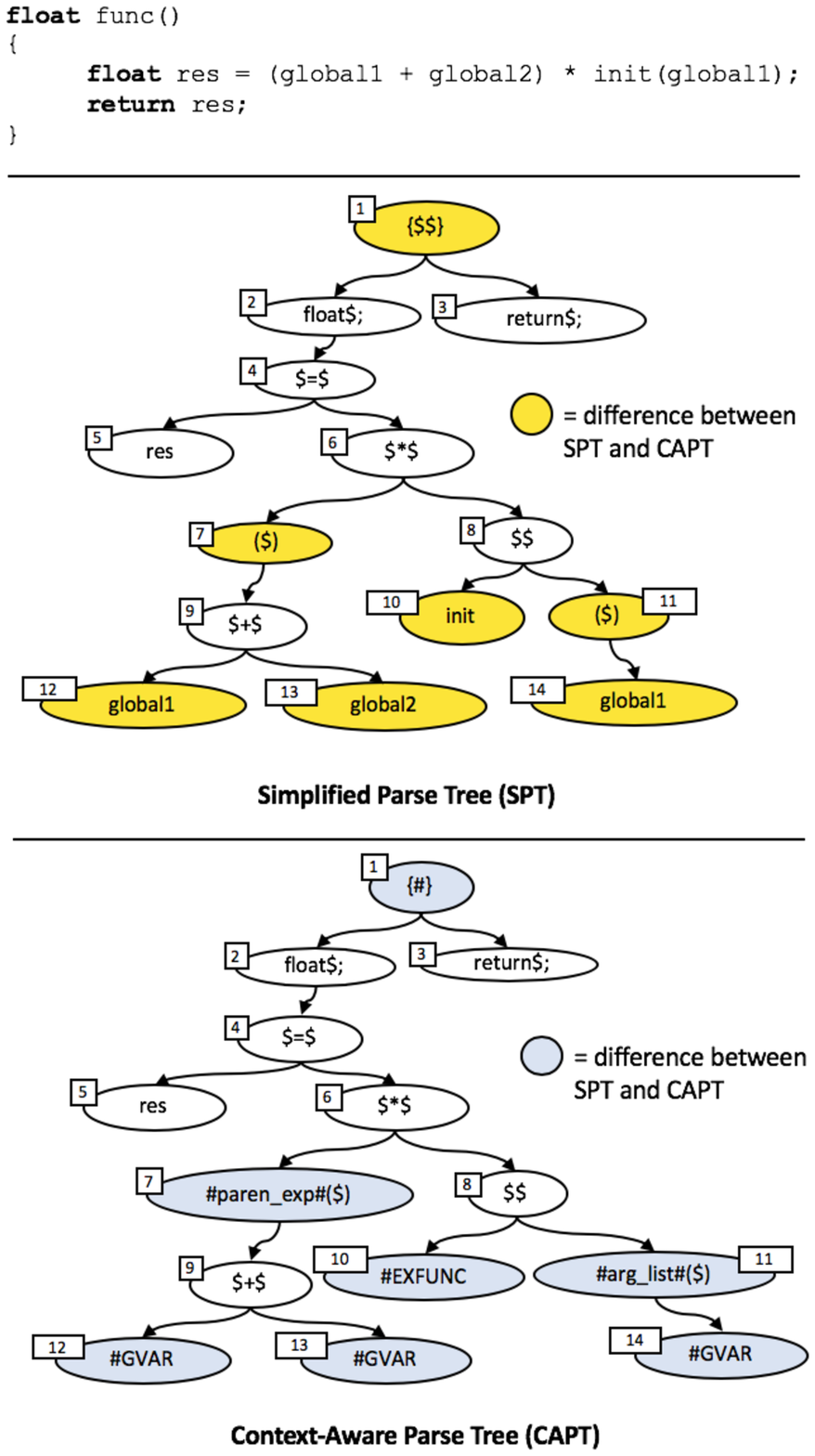}
\caption{Example of the Differences in the SPT and CAPT.}
\label{fig:spt_capt_tree}
\end{center}
\end{figure}

In this work, we present a new structural representation, a context-aware parse tree (CAPT). In contrast to syntactic representations, such as an abstract syntax tree (AST), CAPT is designed to capture the semantic meaning of the user's code. That is, CAPT provides information relevant to whether two code snippets~\footnote{For the purposes of this work, we precisely define \emph{code snippet} as a C/C++ function, discussed in more detail in Section~\ref{sect:capt}.} are semantically convergent, even if they are syntactically divergent. \emph{Code similarity} is informally defined as the process of determining whether two code snippets are semantically similar. In this work, we show that CAPTs are competitive with simplified parse trees (SPTs), the representation used by the Aroma~\cite{luan:2019:oopsla} state-of-the-art code recommendation system. From the point of view of machine programming~\cite{gottschlich:2018:mapl}, this work seeks to improve our ability to automatically recognize a user's intent from code. Thus, it principally falls in the "intention pillar."~\cite{gottschlich:2018:mapl} In addition, once such intention is understood it may be augmented or transferred from language to language, thereby advancing the ``adaptation pillar.''

\paragraph{Why Not An AST?}
While the abstract syntax tree (AST) has had tremendous, demonstrable value in cases where syntactic structure is of primary importance (e.g., source code compilation~\cite{aho:2006:dragon}), the utility of the AST in the space of extracting semantic meaning from code (i.e., lifting intention~\cite{kamil:2016:pldi, ginsbach:2018:asplos, ahmad:2019:siggraph}) may be less clear. The AST contains a complete syntactic representation of the program, which can contain many details relevant to program compilation, but that may be less salient for semantic analysis. For example, an AST for a single line of C/C++ code like, \[ \texttt{int x = (y+3);} \] may contain separate nodes for a variable declaration, a compound statement, and three implicit casts.\footnote{https://clang.llvm.org/docs/IntroductionToTheClangAST.html\#examining-the-ast} While such details are critical for correctly implementing a compiler, they may not be relevant for semantic analysis.
For example, \[ \texttt{int x = y;} \] may be considered semantically similar to the prior code snippet, even if the second operation does not have implicit casts or compound statements. Unlike an AST, a CAPT can omit these details to help improve similarity analysis. We note that the Aroma authors have previously illustrated some of these AST limitations as well~\cite{luan:2019:oopsla}. 

\paragraph{Why Not An SPT?} The Aroma team~\cite{luan:2019:oopsla}, inspired, at least partially, by weaknesses in the AST (as stated directly by the authors), introduced the simplified parse tree (SPT). An SPT is a new structural representation for code similarity, which intentionally departs from the AST. By design, the SPT reduces the syntactic information collected from source code: each node in the SPT is strictly a token from the original program (no other nodes are introduced). The Aroma authors demonstrate that this reduction, or in some cases elimination, of lower-level syntactic information can be helpful for code similarity systems. Such a reduction may be especially salient in the context of type 3 and 4 similarity, where code may be syntactically different, but is semantically similar~\cite{roy:2009:jscico}. 

While the AST may contain too much low-level syntactic information, the SPT may omit semantically-relevant information. For example, because the SPT contains only tokens, the SPT effectively omits whether or not a token binds to a local or a global variable. On the other hand, the SPT may also include too much specificity. For example, the exact name of a variable or function is not always relevant to its semantic meaning (e.g., two global variables with the same name but from different programs might not imply similar semantics). In light of this, we identified two areas where an SPT can be modified such that the resulting tree yields improved code similarity accuracy. Those areas are: \emph{(i)} language-agnostic modifications, which remove potentially irrelevant syntactic information, and \emph{(ii)} language-specific modifications, which introduces new syntactic information that may be semantically salient. We believe CAPTs can capture the essence of these two design elements. Our paper provides the following technical contributions:

\begin{itemize}

    \item We introduce the \emph{context-aware parse tree} (CAPT), a novel modification of the simplified parse tree (SPT) intended to improve code similarity analysis.

    \item We illustrate and discuss the two flexibility enhancements CAPT has compared to Aroma's simplified parse tree: \emph{(i)} language-agnostic and \emph{(ii)} language-specific.

    \item Our research quantitatively demonstrates the value of our proposed semantically-salient features, enabling a specific CAPT configuration to be 39\% more accurate than SPT across the 48,610 programs we analyzed.

\end{itemize}



%% file: capt.tex
\section{System Design}
\begin{figure}[htpb]
\begin{center}
\includegraphics[width=0.75\columnwidth]{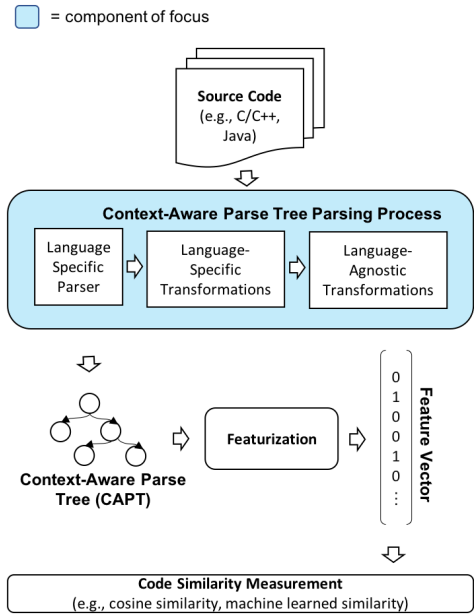}
\vspace{-0.25cm}
\caption{The MISIM Code Similarity System.}
\label{fig:system_overview}
\end{center}
\end{figure}

Before discussing the specifics of our approach, we first provide some background on how both CAPTs and SPTs are used in code similarity systems. Figure~\ref{fig:system_overview} presents an abbreviated overview of our code similarity system, MISIM (we illustrate MISIM using CAPTs, but any process that transforms a code snippet into a feature vector could be used, including SPTs). Figure~\ref{fig:system_overview} illustrates the process of transforming source code (e.g., a C function) to a feature vector. Once a feature vector is generated, a code similarity measurement (e.g., vector dot product~\cite{lipschutz:1968:mcgraw-hill}, cosine similarity~\cite{singhal:2001:ieee}, machine-learned similarity~\cite{zhao:2018:esec/fse}) calculates the similarity score between the input program and other programs stored in a database.
~\footnote{In the context of this work, we perform code similarity analysis on an entire C/C++ program, where we differentiate code snippets by uniquely defined C/C++ function bodies.} Although the current system only supports C/C++ code, our design is agnostic to the underlying programming language. While we have built a prototype of the entire system, CAPT is the emphasis of this paper. As such, we omit a deeper dissection of other components of the system as we consider them outside of the scope of this paper.

To generate a CAPT, the code is first parsed into a language-agnostic parse tree. Next, the system performs language-specific transformations in constructing an initial intermediate form of CAPT by adding pertinent  information used to disambiguate code. The system then performs language-agnostic transformations (e.g., abstracting the number of code statements in a function) by potentially pruning or modifying the CAPT's nodes. Subsequently, a CAPT is featurized into a vector using the same procedure as SPT's featurization process~\cite{luan:2019:oopsla}. 

\section{The Context-Aware Parse Tree (CAPT)}
\label{sect:capt}
In this section, we describe the fundamental design of CAPT, including the key differences between CAPT and Aroma's simplified parse tree (SPT). Some of these details are visually illustrated in Figure~\ref{fig:spt_capt_tree}. Fundamentally, a CAPT is the result of transforming an SPT in specific ways according to a \emph{configuration}, options that give CAPT a greater degree of flexibility. Different configurations may result in better performance in some domains, but worse performance in others. Here, we focus on describing the intuition behind these options, and evaluating every possible configuration in one domain. In this preliminary work, we do not address figuring out which configuration to use in a particular domain, but plan to investigate this in future work.

\paragraph{CAPT Configuration Categories (see Table~\ref{tab:configurations})}


CAPT's configuration categories come in two general forms: \emph{language-specific} configurations and \emph{language-agnostic} configurations, listed in Table~\ref{tab:configurations}. We next give an intuitive overview of both.

Table~\ref{tab:configurations} lists the current types and options for the language-specific and language-agnostic categories in CAPT.~\footnote{While our exploration into CAPT is still early, we believe our categories may be exhaustive (that is, fully encompassing). Yet, we do not believe our configuration types or options are exhaustive.} Each configuration type has multiple options associated with it to afford the user the flexibility in exploring a number of CAPT configurations. For all configuration types, option 0 always corresponds to the Aroma system's original SPT. Each of the types in Table~\ref{tab:configurations} are described in greater detail in the following sections. 

Language-specific configurations, described in Section~\ref{subsec:language-specific} are designed to resolve syntactic ambiguity present in the SPT. For example, in Figure~\ref{fig:spt_capt_tree}, the SPT treats the parenthetical expression \texttt{(global1 + global2)} identically to the parenthetical expression \texttt{init(global1)}, whereas the CAPT configuration shown disambiguates these two terms (the first is a parenthetical expression, the second is an argument list). Such a disambiguation may be useful to a code similarity system, as the CAPT representation makes the presence of a function call more clear. 

Language-agnostic configurations, described in Section~\ref{subsec:language-agnostic}, can improve code similarity analysis by unbinding overly-specific semantics that may be present in the original SPT structure. For example, in Figure~\ref{fig:spt_capt_tree}, the SPT includes the literal names \texttt{global1}, \texttt{global2}, etc. The CAPT variant, on the other hand, unbinds these names and replaces them with a generic string (\texttt{\#GVAR}). This could improve code similarity analysis if the exact token names are irrelevant, and the semantically-salient feature is simply that there is a global variable.

We note that that these examples are not universal. One specific CAPT configuration is unlikely to work in all scenarios: sometimes, disambiguating parenthetical expressions may be good, other times, it may be bad. This work seeks to explore and analyze these possible configurations. We provide a formalization and concrete examples of both language-agnostic and language-specific configurations later in this section.

\begin{table}[]
\centering
\caption{CAPT Configuration Options.}
\vspace{-3mm}
\label{tab:configurations}
\resizebox{\columnwidth}{!}{%
\begin{tabular}{|l|l|l|}
\hline
Category & Type                                 & Option                                                  \\\hline
\multirow{3}{*}{Language-specific}  & \multirow{3}{*}{A. Node Annotations}    & 0. No change (Aroma's original configuration)  \\
         &                                      & 1. Annotate all internal nodes                          \\
         &                                      & 2. Annotate parenthesis nodes (C/C++ Specific)                           \\\hline
\multirow{10}{*}{Language-agnostic} & \multirow{3}{*}{B. Compound Statements} & 0. No change (Aroma's original configuration) \\
         &                                      & 1. Remove all features relevant to compound statements  \\
         &                                      & 2. Replace with `\{\#\}'                                \\\cline{2-3}
         & \multirow{4}{*}{C. Global Variables} & 0. No change (Aroma's original configuration)                                       \\
         &                                      & 1. Remove all features relevant to global variables     \\
         &                                      & 2. Replace with `\#GVAR'                                \\
         &                                      & 3. Replace with `\#VAR' (the label for local variables) \\\cline{2-3}
         & \multirow{3}{*}{D. Global Functions} & 0. No change (Aroma's original configuration)                                       \\
         &                                      & 1. Remove all features relevant to global functions     \\
         &                                      & 2. Replace with `\#EXFUNC'                              \\\hline
\end{tabular}%
}
\vspace{-3mm}
\end{table}

\begin{figure}[htpb]
\begin{center}
\includegraphics[width=0.75\columnwidth]{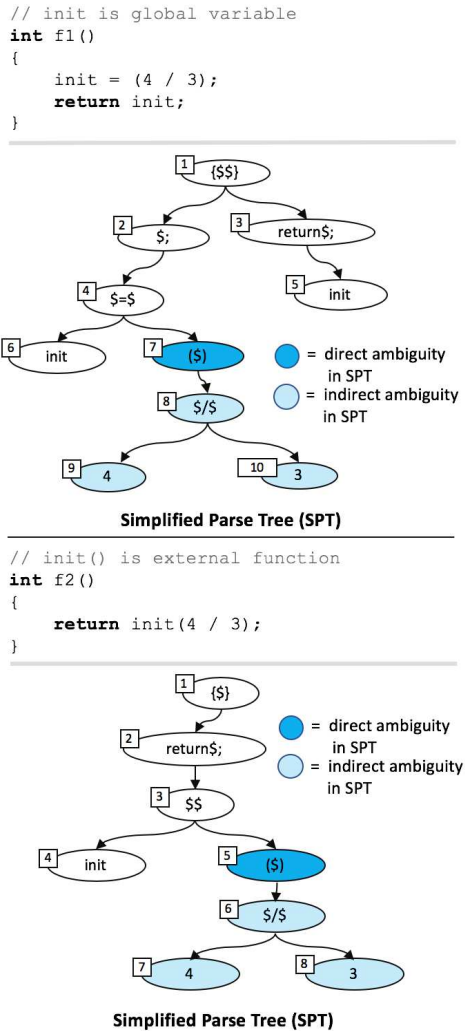}
\vspace{-0.2cm}
\caption{Language Ambiguity in Simplified Parse Tree.}
\label{fig:spt_ambiguity}
\end{center}
\vspace{-0.25cm}
\end{figure}

\subsection{Language-Specific Configurations}
\label{subsec:language-specific}

Language-specific configurations are meant to capture semantic meaning by resolving ambiguity and introducing specificity related to the specific underlying programming language. Intuitively, these configurations can be thought of as \emph{syntax-binding}, capturing semantic information that are bound to the particular syntactical structure of the program. In some cases, these specifications may capture relevant semantic information, whereas in other cases these specifications may capture irrelevant details.


\paragraph{Node Annotations.}

We define a \emph{node annotation} as a modification to a tree's node to incorporate more information. Node annotations are generally used to facilitate disambiguation caused by language-specific syntax ambiguity. These ambiguous scenarios tend to arise when certain code constructs and/or operators have been overloaded in a specific language. In such cases, the original SPT structure may be insufficient to properly disambiguate between them, potentially reducing its ability to evaluate code semantic similarity (see Figure~\ref{fig:spt_ambiguity}). CAPT’s node annotation options are meant to help resolve this. 
As we incorporate more language-specific syntax into CAPT nodes, we run the chance of overloading the tree with syntactic details. This could potentially undo the general reasoning behind Aroma's SPT and our CAPT structure. We discuss this in greater detail in Section~\ref{ssect:weakness}.


\subsubsection{C and C++ Node Annotations}

For our first embodiment of CAPT, we have focused solely on C and C++ programs. We have found that programs in C/C++ present at least two interesting challenges.\footnote{We do not claim that these challenges are \emph{unique} to C/++: these challenges may be present in other languages as well.}

\paragraph{(Lack of) Compilation.} We have found that, unlike programs in higher-level programming languages (e.g., Python~\cite{rossum:2009:python3}, JavaScript~\cite{flanagan:2006:javascript}), C/C++ programs found "in the wild" tend to not immediately compile from a source repository (e.g., GitHub~\cite{cosentino:2017:ieee}). Thus, code similarity analysis may need to be performed without relying on successful compilation.

\paragraph{Many Solutions.} The C and C++ programming languages provide multiple diverse ways to solve the same problem (e.g., searching a list with a for loop vs. using \texttt{std::find}). Because of this, C/C++ enables programmers to create semantically similar programs that are syntactically divergent. In extreme cases, such semantically similar (or identical) solutions may differ in computation time by orders of magnitude~\cite{satish:2012:isca}. This requires that code similarity techniques to be robust in their ability to identify semantic similarity in the presence of syntactic dissimilarity (i.e. a type 4 code similarity exercise~\cite{roy:2009:jscico}).

\noindent
\\
We believe that analytically deriving the optimal selection of node annotations across all C/C++ code may be untenable. To accommodate this, we currently provide two levels of granularity for C/C++ node annotations in CAPT.~\footnote{This is still early work and we expect to identify further refinement options in C/C++ and other languages as the research progresses.} 
\begin{itemize}
\item{\textbf{Option 0}: original Aroma SPT configuration.}
\item{\textbf{Option 1}: annotation of all nodes with their language-specific node type.}
\item{\textbf{Option 2}: annotation of all nodes containing parentheticals with their language-specific node type.}
\end{itemize}

Option 1 corresponds to an extreme case of a concrete syntax embedding (e.g., every node contains syntactic information, and all syntactic information is represented in some node). Since such an embedding may "overload" the code similarity system with irrelevant syntactic details, Option 2 can be used to annotate only parentheticals, which we have empirically identified to often have notably divergent semantic meaning based on context. 

An example is shown in Figure~\ref{fig:spt_capt_tree}. In one case the parentheses is applied as a mathematical precedence operator, in the other it is used as a function call. If left unresolved, such ambiguity would cause the subtree rooted at node 7 of function \texttt{f1} to be classified identically to the subtree rooted at node 5 of function \texttt{f2}. The intended purpose of the parenthesis operator is context sensitive and is disambiguated by encoding the contextual information into the two distinct node annotations, i.e. the parenthesized expression and the argument list respectively. 

\subsection{Language-Agnostic Configurations}
\label{subsec:language-agnostic}

Unlike language-specific configurations, language-agnostic configurations are not meant to be restricted to the specific syntax of a specific language. Instead, they are meant to be applied generally across multiple languages. Intuitively, these configurations can be thought of as \emph{syntax-unbinding} in nature: they generally abstract (or, in some cases, entirely eliminate) syntactical information in the attempt to improve its ability to derive semantic meaning from the code.

\paragraph{Compound Statements.} 
The \textit{compound statements} configuration is a language-agnostic option that enables the user to control how much non-terminal node information is incorporated into the CAPT. Again, Option 0 corresponds to the original Aroma SPT. Option 1 omits separate features for compound statements altogether. Option 2 does not discriminate between compound statements of different lengths and specifies a special label to denote the presence of a compound statement. For example, the \textit{for} loop construct in C/C++ is represented with a single label with this option instead of constructing three separate labels for the loop initialization, test condition and increment.

\paragraph{Global Variables.}
The \textit{global variables} configuration specifies the degree of global variable-specific information contained in a CAPT. In addition to Aroma's original configuration (Option 0), which annotates nodes by including the precise global variable name, CAPT provides three additional configurations. Option 1 specifies the extreme case of eliding all information on global variables. Option 2 annotates all global variables with the special label ‘\#GVAR’, omitting the names of the global variable identifiers. Option 3 designates global variables with the label ‘\#VAR’ rendering them indistinguishable from the usage of local variables. 

Intuitively, including the precise global variable names (Option 0) may be appropriate if code similarity is being performed on a single code-base, where two references to a global variable with the same name necessarily refer to the same global variable. Options 1 through 3, which remove global variable information to varying degrees, may be appropriate when performing code similarity between unrelated code-bases, where two different global variables named (for example) \texttt{foo} are most likely unrelated.

\paragraph{Global Functions.}
The \textit{global functions} configuration serves the dual purpose of \emph{(i)} controlling the amount of function-specific information to featurize and \emph{(ii)} to disambiguate between the usage of global functions and global variables in CAPT, a feature that is curiously absent in the original SPT design: the SPT shown in Figure~\ref{fig:spt_capt_tree} has no distinction between \texttt{init} (a function) and \texttt{global1} (a variable). Option 1 removes all features pertaining to global functions. Option 2 annotates all global function references with the special label ‘\#EXFUNC’ while eliminating the function identifier. Intuitively, these options behave similarly to the global variable options. Our current prototype, which handles only single C/C++ functions, does not differentiate between external functions. In future work, we plan to investigate CAPT variants that differentiate between local, global, and library functions.


\subsection{Discussion}
\label{subsec:intuit}
We believe there is no silver bullet solution for code similarity for all programs and programming languages. Based on this belief, a key intuition of CAPT's design is to provide a structure that is semantically rich based on structure, with heavy inspiration from Aroma's SPT, while simultaneously providing a range of customizable parameters to accommodate a wide variety of scenarios. CAPT's language-agnostic and language-specific configurations and their associated options serve for exploration of a series of tree variants, each differing in their granularity of detail of abstractions. 

For instance, the \textit{compound statements} configuration provides three levels of abstraction. Option 0 is Aroma's baseline configuration and is the finest level of abstraction, as it featurizes the number of constituents in a compound statement node. Option 2 reduces compound statements to a single token and represents a slightly higher level of abstraction. Option 1 eliminates all features related to compound statements and is the coarsest level of abstraction. The same trend applies to the \textit{global variables} and \textit{global functions} configurations. It is our belief, based on early evidence, that the appropriate level of abstraction in CAPT is likely based on many factors such as \emph{(i)} code similarity purpose, \emph{(ii)} programming language expressiveness, and \emph{(iii)} application domain.

Aroma's original SPT seems to work well for a common code base where global variables have consistent semantics and global functions are standard API calls also with consistent semantics (e.g., a single code-base). However, for cases outside of such spaces, some question about applicability arise. For example, assumptions about consistent semantics for global variables and functions may not hold in cases of non-common code-bases or non-standardized global function names~\cite{wulf:1973:sigplan, gellenbeck:1991:ablex, feitelson:2020:ieee}. The capacity to differentiate between these cases, and others, is a key motivation for CAPT.

We do not believe that CAPT's current structure is exhaustive. With this in mind, we have designed CAPT to be extensible, enabling a seamless mechanism to add new configurations and options (described in Section~\ref{sect:experiments}). Our intention with this paper is to present initial findings in exploring CAPT's structure. Based on our early experimental analysis, presented in Section~\ref{sect:experiments}, CAPT seems to be a promising research direction for code similarity.

\paragraph{An Important Weakness}
\label{ssect:weakness}

While CAPT provides added flexibility over SPT, such flexibility may be misused. With CAPT, system developers are free to add or remove as much syntactic differentiation detail they choose for a given language or given code body. Such overspecification (or  underspecification), may result in syntactic overload (or underload) which may cause reduced code similarity accuracy over the original SPT design, as we illustrate in Section~\ref{sect:experiments}.

%% file: experiments.tex
\section{Experimental Results}
\label{sect:experiments}

In this section, we discuss our experimental setup and analyze the performance of CAPT compared to Aroma's simplified parse tree (SPT). In Section~\ref{subsec:dataset}, we describe the code corpus used with CAPT that includes hundreds of unique programming solutions for 104 different programming problems. In Section~\ref{subsec:setup}, we explain the dataset grouping and enumeration for our experiments. We also discuss the metrics used to quantitatively rank the different CAPT configurations and those chosen for evaluation of code similarity. Section~\ref{subsec:results} demonstrates that, a code similarity system built using CAPT  \emph{(i)} has a greater frequency of improved accuracy for the total number of problems and \emph{(ii)} is, on average, more accurate than SPT. For completeness, we also include cases where CAPT configurations perform poorly.


\subsection{Dataset}
\label{subsec:dataset}

Our experiments use the POJ-104 dataset. The POJ-104 dataset is the result of educationally-inspired programming questions, which consist of student written programs to 104 problems~\cite{mou:2016:aaai}. Each problem has 500 unique solutions written in C/C++. Each solution has been validated for correctness. We categorize all solutions for a given POJ-104 problem as being in the same semantic similarity equivalence class. We make no claims about the semantic similarity or dissimilarity of solutions to two or more different POJ-104 problems. 

Using this approach, we treat the problem of code similarity analysis as a classification problem. We classify two programs as semantically similar if they originate from the same equivalence class (i.e., the same POJ-104 problem). Using this approach, the labels for these classifications can be implicitly lifted using the problem's unique identifier.

\paragraph{Eliminated Programs From POJ-104.}
Some of the coding solutions in the POJ-104 dataset have been marked as illegal by the parser we used~\footnote{Tree-sitter: \url{http://tree-sitter.github.io/tree-sitter}}. After investigating, we found this to be due to the code using non-standard coding conventions (e.g., unspecified return types, lack of semicolons at the end of structure definitions, etc.). Because they could not be properly parsed, we have pruned them from the dataset. We also eliminated all of the solutions that we could find that had hard-coded answers to problems. \footnote{Unfortunately, due to the size of the dataset, we cannot guarantee all such programs were eliminated.} The resulting dataset consists of 48,610 programming solutions with 370 to 499 uniquely coded solutions per problem.

\begin{figure*}[h]
\centering
\begin{subfigure}[t]{0.33\linewidth}
\captionsetup[subfigure]{width=0.9\textwidth}
\captionsetup{justification=centering}
\includegraphics[width=\linewidth]{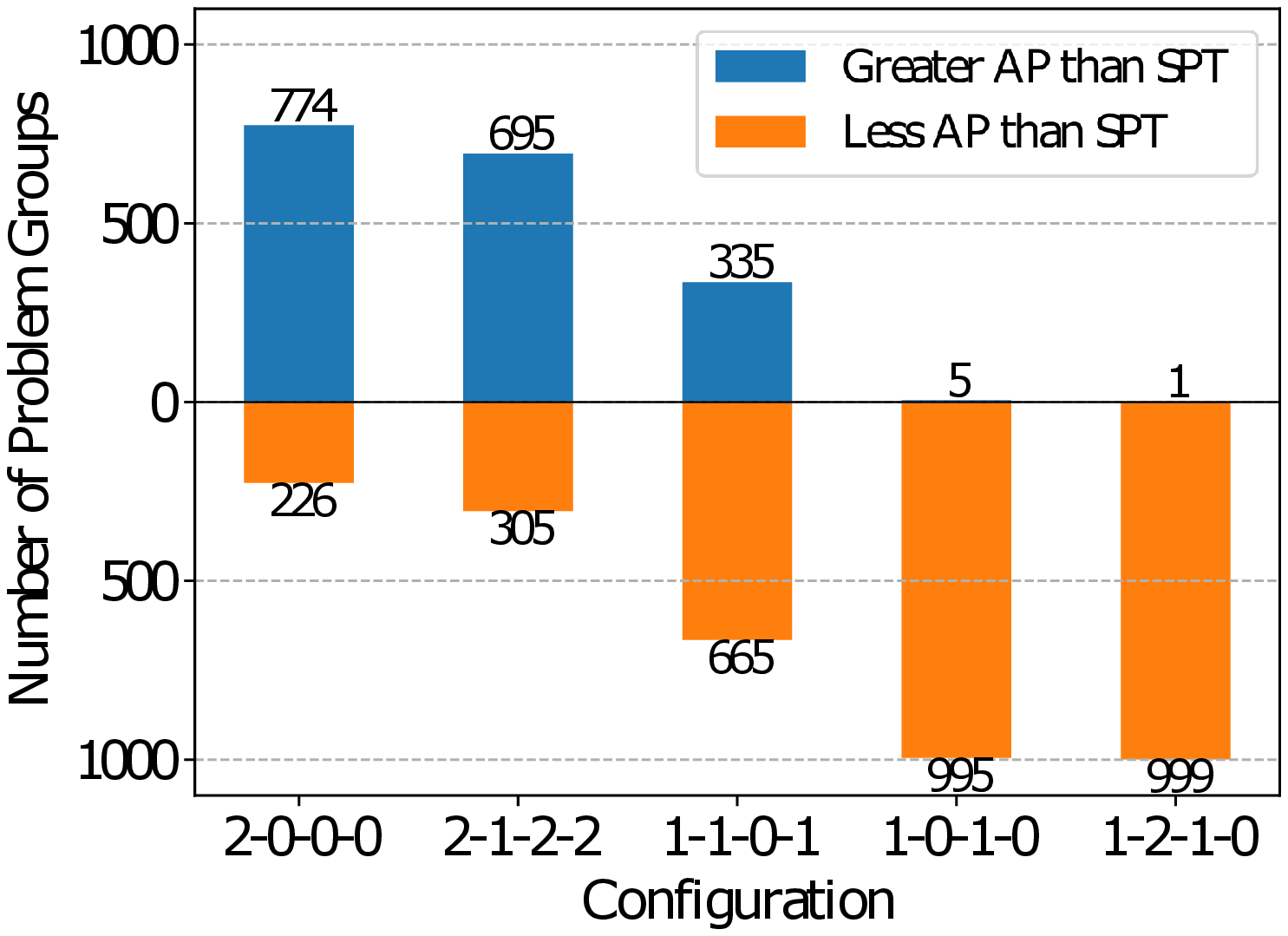}
\caption{Breakdown of the Number of Groups with AP Greater or Less than SPT.}
\label{fig:ap_gl}
\end{subfigure}
\begin{subfigure}[t]{0.33\linewidth}
\captionsetup[subfigure]{width=0.9\textwidth}
\captionsetup{justification=centering}
\includegraphics[width=\linewidth]{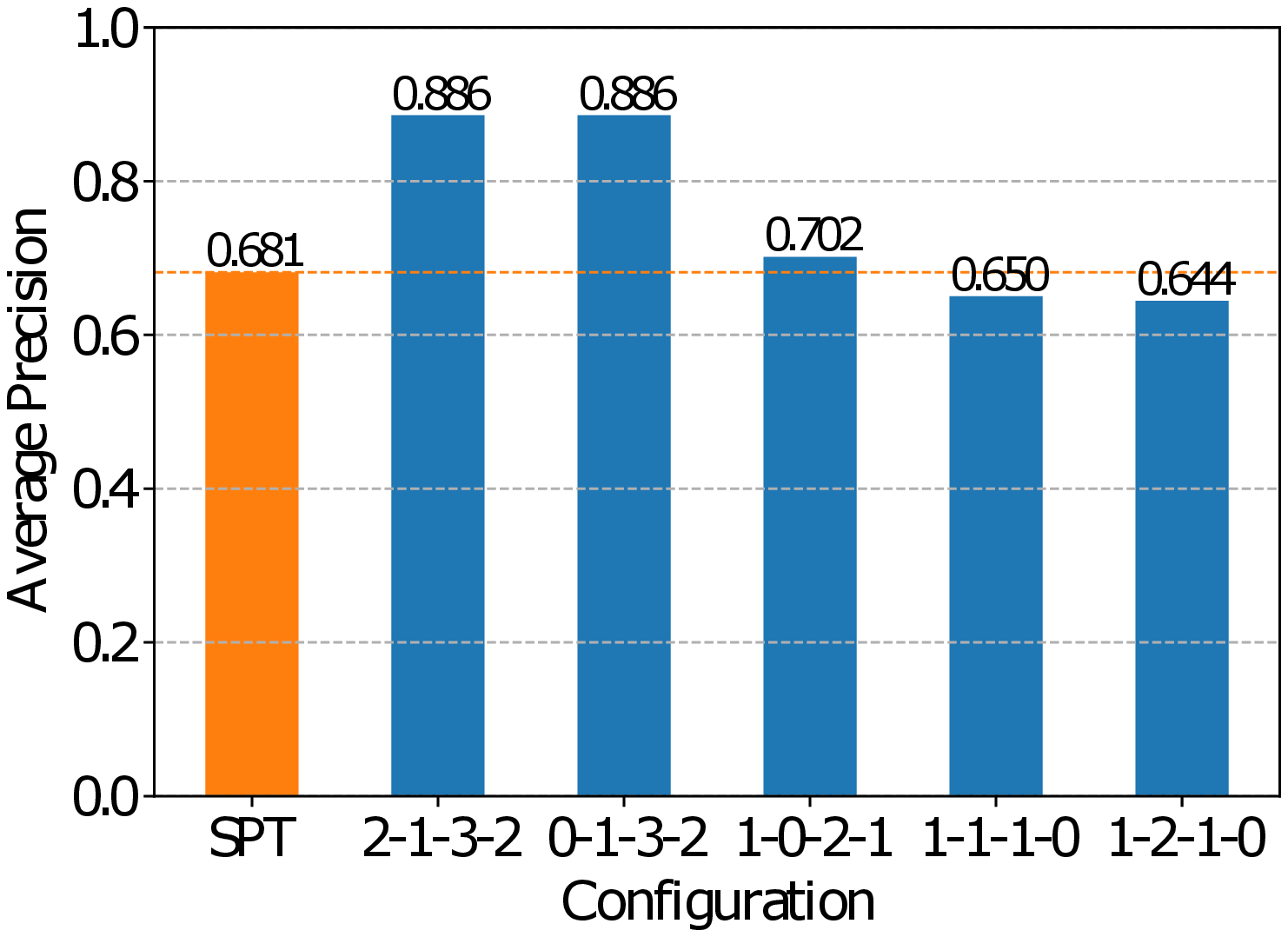}
\caption{Average Precision for the Group Containing the Best Case.}
\label{fig:best_group}
\end{subfigure}
\begin{subfigure}[t]{0.33\linewidth}
\captionsetup[subfigure]{width=0.9\textwidth}
\captionsetup{justification=centering}
\includegraphics[width=\linewidth]{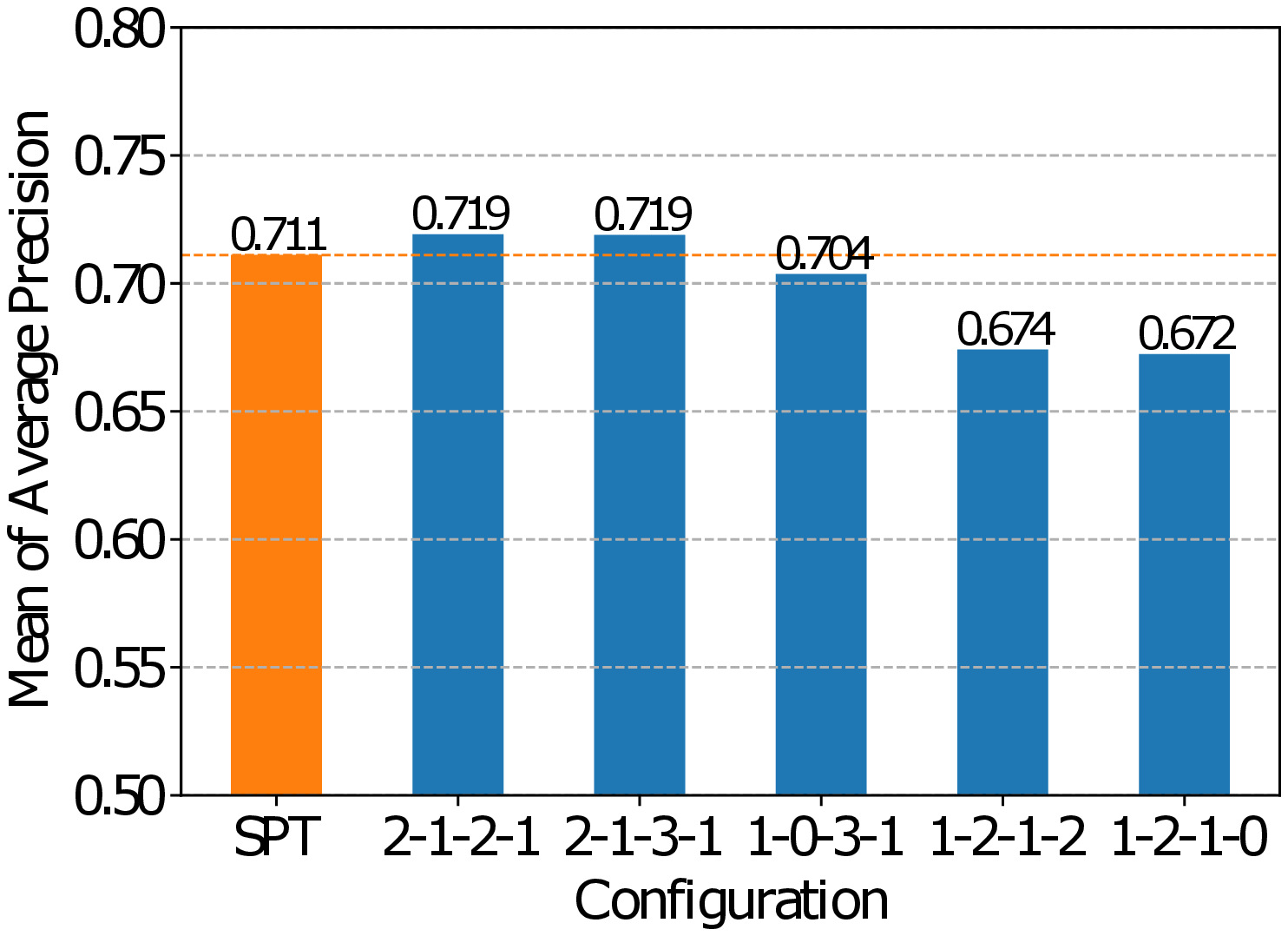}
\caption{Mean of Average Precision Over All Program Groups.}
\label{fig:ap_mean}
\end{subfigure}
\caption{Comparison of CAPT and SPT. The blue bars in \textbf{(a)} and \textbf{(b)}, and all the bars in \textbf{(c)}, from left to right, correspond to the best two, the median, and the worst two CAPT configurations, ranked by the metric displayed in each subfigure.}
\label{fig:ap_all}
\end{figure*}


\subsection{Experimental Setup}
\label{subsec:setup}
In this section, we describe our experimental setup. At the highest level, we compare the performance of various configurations of CAPT to Aroma's SPT. The list of possible CAPT configurations are shown in Table~\ref{tab:configurations}.

\paragraph{Problem Group Selection.}
Given that POJ-104 consists of 104 unique problems and 48,610 programs, depending on how we analyze the data, we might face intractability problems in both computational and combinatorial complexity. With this in mind, our initial approach is to construct 1000 sets of five unique, pseudo-randomly selected problems for code similarity analysis. Using this approach, we evaluate every configuration of CAPT and Aroma's original SPT on each pair of solutions for each problem set. We then aggregate the results across all the groups to estimate their overall performance. While this approach is not exhaustive of possible combinations (in set size or set combinations), we aim for it to be a reasonable starting point. As our research with CAPT matures, we plan to explore a broader variety of set sizes and a more exhaustive number of combinations.

\paragraph{Code Similarity Performance Evaluation.}
For each problem group, we exhaustively calculate code similarity scores for all unique solution pairs, including pairs constructed from the same program solution (i.e., program $A$ compared to program $A$). We use $G$ to refer to the set of groups and $g$ to indicate a particular group in $G$. We denote $|G|$ as the number of groups in $G$ (i.e. cardinality) and |g| as the number of solutions in group $g$. For $g$ = $G_i$, where $i = \{1, 2, \ldots, 1000\}$, the total unique program pairs (denoted by $g_P$) in $G_i$ is $\mathit{|g_{P}|} = \frac{1}{2}|g|(|g|+1)$.

To compute the similarity score of a solution pair, we use Aroma's approach. This includes calculating the dot product of two feature vectors (i.e., a program pair), each of which is generated from a CAPT or SPT structure. The larger the magnitude of the dot product, the greater the similarity.

We evaluate the quality of the recommendation based on \emph{average precision}. \emph{Precision} is the ratio of true positives to the sum of true positives and false positives. Here, true positives denote solution pairs correctly classified as similar and false positives refer to solution pairs incorrectly classified as similar. \emph{Recall} is the ratio of true positives to the sum of true positives and false negatives, where false negatives are solution pairs incorrectly classified as different. As we monotonically increase the threshold from the minimum value to the maximum value, precision generally increases while recall generally decreases. The \textit{average precision} (AP) summarizes the performance of a binary classifier under different thresholds for categorizing whether the solutions are from the same equivalence class (i.e., the same POJ-104 problem)~\cite{liu:2009:now}. AP is calculated using the following formula over all thresholds.

\begin{enumerate}
    \item All unique values from the $M$ similarity scores, corresponding to the solution pairs, are gathered and sorted in descending order. Let $N$ be the number of unique scores and $s_1, s_2, \ldots, s_N$ be the sorted list of such scores.
    
    \item For $i$ in $\{1, 2, \ldots, N\}$, the precision $p_i$ and recall $r_i$ for the classifier with the threshold being $s_i$ is computed.
    \item Let $r_0 = 0$. The average precision is computed as: $$AP = \sum_{i=1}^N(r_i-r_{i-1})p_i$$
\end{enumerate}

\paragraph{Configuration Identifier.}
In the following sections, we refer to a configuration of CAPT by its unique identifier (ID). A configuration ID is formatted as A-B-C-D. Each of the four letters corresponds to a configuration type in the second column of Table~\ref{tab:configurations}, and will be replaced by an option number specified in the third column of the table. Configuration 0-0-0-0 corresponds to Aroma's SPT.


\subsection{Results}
\label{subsec:results}

Figure~\ref{fig:ap_gl} depicts the number of problem groups where a particular CAPT variant performed better (blue) or worse (orange) than SPT. For example, the CAPT configuration 2-0-0-0 outperformed SPT in 774 of 1000 problem groups, and underperformed in 226 problem groups. This equates to a 54.8\% accuracy improvement of CAPT over SPT. Figure~\ref{fig:ap_gl} shows the two best (2-0-0-0 and 2-1-2-2), the median (1-1-0-1), and the two worst (1-0-1-0 and 1-2-1-0) configurations with respect to SPT. 2-1-2-2 demonstrates a CAPT configuration where all its options are exercised and not tunable in SPT. This configuration performs better than SPT on 695 of the 1000 problem groups, i.e. on $\approx39\%$ of the problem groups. Although it is less performant than the 2-0-0-0 configuration, it exercises all of CAPT's unique tunable parameters. We speculate that these configuration results may vary based on programming language, domain, and problem type, amongst other parameters.

Figure~\ref{fig:best_group} shows the group containing the problems for which CAPT achieved the best performance relative to SPT, among all 1000 problem groups. In other words, Figure~\ref{fig:best_group} shows the performance of SPT and CAPT for the single problem group with the greatest difference between a CAPT configuration and SPT. In this single group, CAPT achieves the maximum improvement of more than 30\% over SPT for this problem group on two of its configurations. We note that, since we tested 108 CAPT configurations across 1000 different problem groups, there is a reasonable chance of observing such a large difference \emph{even if CAPT performed identically to SPT in expectation.} We do not intend for this result to demonstrate statistical significance, but simply to illustrate the outcome of our experiments. 

Figure~\ref{fig:ap_mean} compares the mean of AP over all 1000 problem groups. In it, the blue bars, moving left to right, depict the CAPT configurations that are \emph{(i)} the two best, \emph{(ii)} the median, and \emph{(iii)} the two worst in terms of average precision. Aroma's baseline SPT configuration is highlighted in orange. The best two CAPT configurations show an average improvement of more than 1\% over SPT, while the others degraded performance relative to the baseline SPT configuration. 

These results illustrate that certain CAPT configurations can outperform the SPT on average by a small margin, and can outperform the SPT on specific problem groups by a large margin. However, we also note that choosing a good CAPT configuration for a domain is essential. We leave automating this configuration selection to future work.

\subsubsection{Analysis of Configurations}

\begin{figure}
\centering
\begin{subfigure}[b]{0.49\columnwidth}
\includegraphics[width=\linewidth]{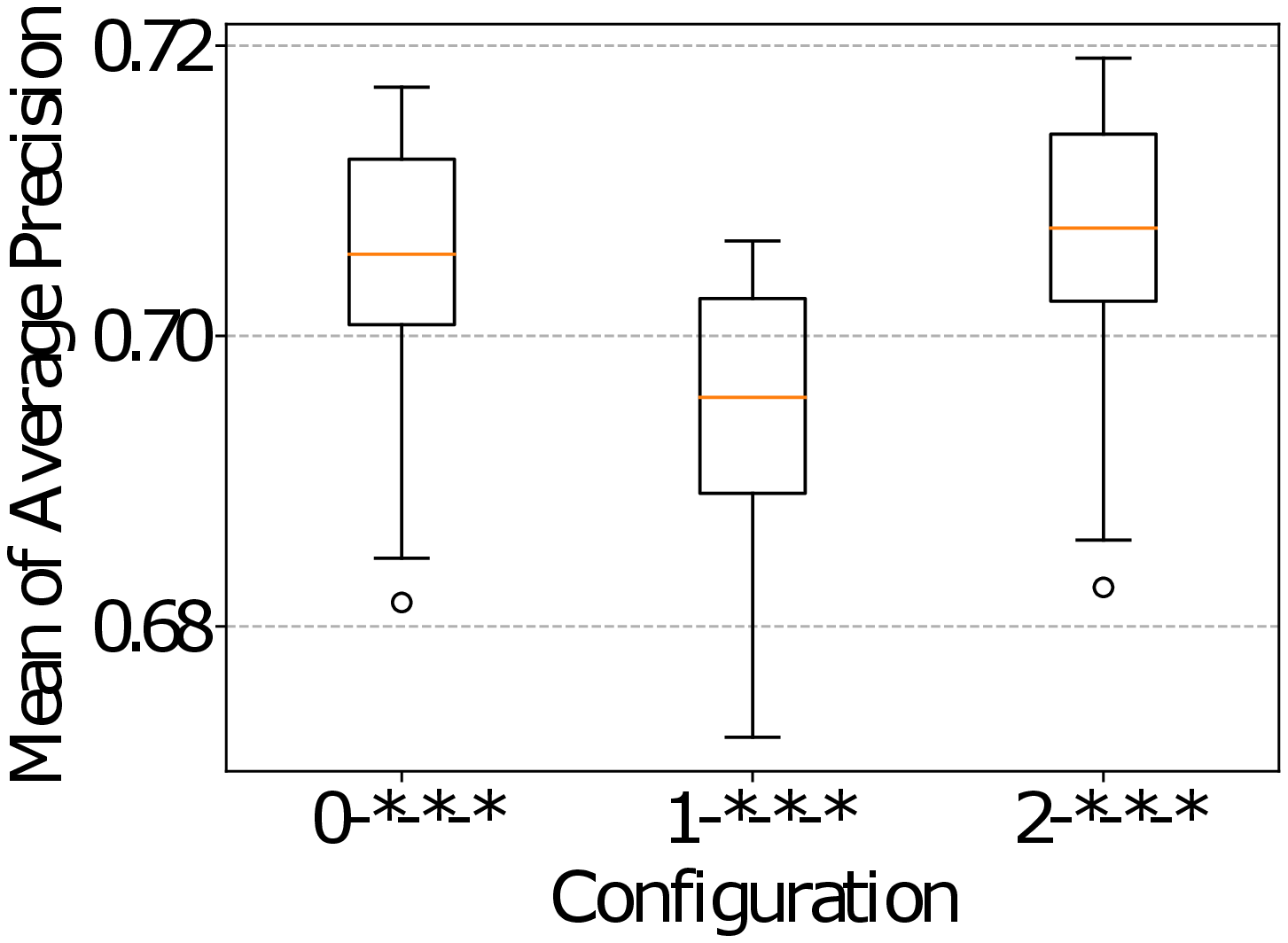}
\caption{Node Annotations.}
\label{fig:c0}
\end{subfigure}
\begin{subfigure}[b]{0.49\columnwidth}
\includegraphics[width=\linewidth]{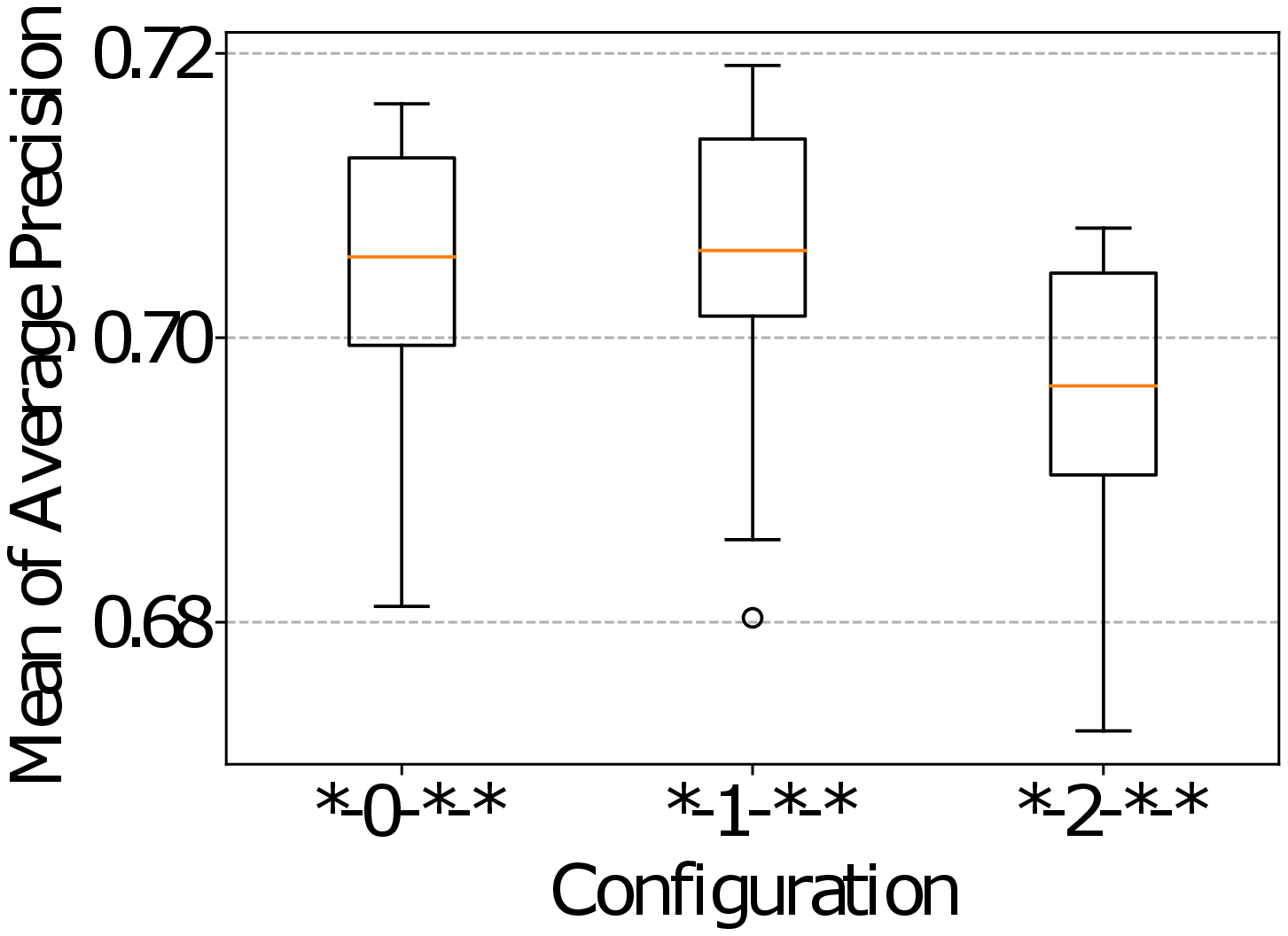}
\caption{Compound Statements.}
\label{fig:c1}
\end{subfigure}
\begin{subfigure}[b]{0.49\columnwidth}
\includegraphics[width=\linewidth]{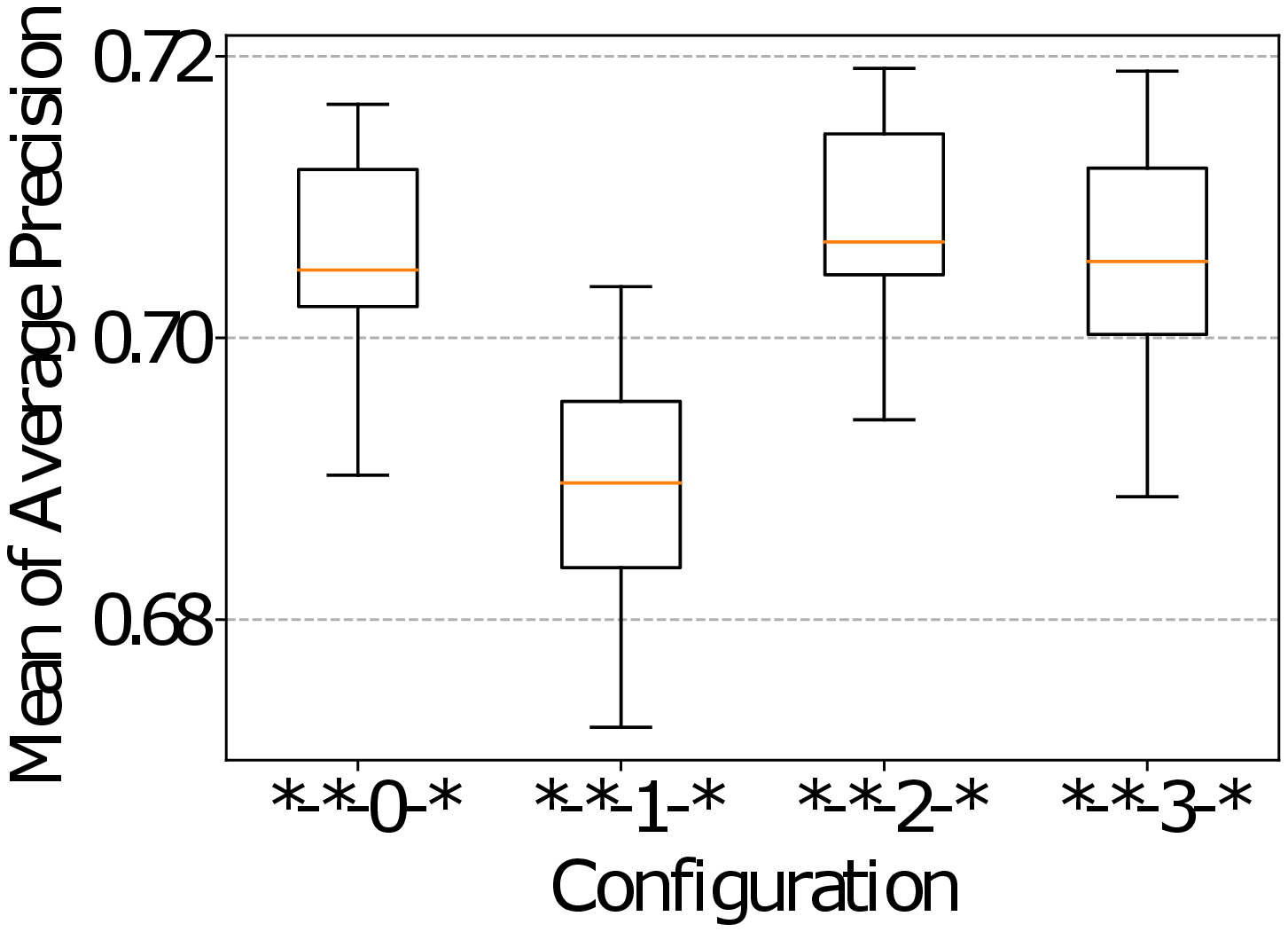}
\caption{Global Variables.}
\label{fig:c2}
\end{subfigure}
\begin{subfigure}[b]{0.49\columnwidth}
\includegraphics[width=\linewidth]{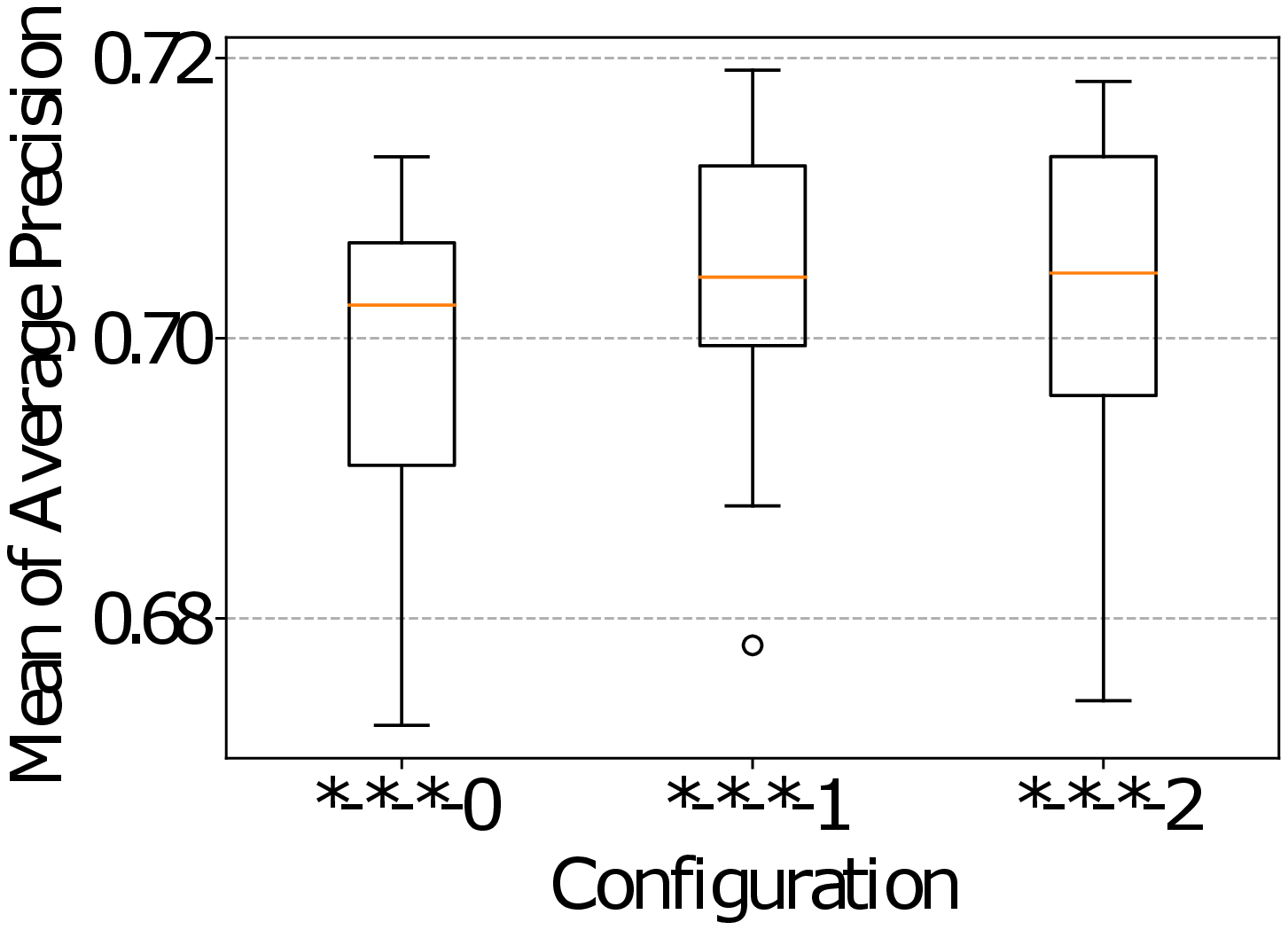}
\caption{Global Functions.}
\label{fig:c3}
\end{subfigure}
\caption{The Distributions of Performance for Configurations with a Fixed Option Type.}
\label{fig:ap_fix_option}
\end{figure}

Figures~\ref{fig:c0}-\ref{fig:c3} serve to illustrate the performance variation for individual configurations. Figure~\ref{fig:c0} shows the effect of varying the options for the \textit{node annotation} configuration. Applying the annotations for the parentheses operator (option 2) results in the best overall performance while annotating every internal node (option 1) results in a concrete syntax tree and the worst overall performance. This underscores the trade-offs in incorporating syntax-binding transformations in CAPT. In Figure~\ref{fig:c1} we observe that removing all features relevant to \textit{compound statements} (option 1) leads to the best overall performance when compared with other options. This indicates that adding separate features for compound statements obscures the code's intended semantics when the constituent statements are also individually featurized. 

Figure~\ref{fig:c2} shows that removing all features relevant to \textit{global variables} (option 1) degrades performance. We also observe that eliminating the global variable identifiers and assigning a tag to signal their presence (option 2) performs best overall, possibly because global variables appearing in similar contexts may not use the same variable identifiers. Further, option 2 performs better than the case where global variables are indistinguishable from local variables (option 3).  Figure~\ref{fig:c3} indicates that removing features relevant to identifiers of \textit{global functions}, but flagging their presence with a special tag as done in option 2, generally gives the best performance. This result is consistent with the intuitions for eliminating features of function identifiers in CAPT as discussed in Section~\ref{subsec:intuit}.

\paragraph{A Subtle Observation.}
A more nuanced and subtle observation is that our results seem to indicate that for each CAPT configuration the optimal granularity of abstraction detail is different. For \textit{compound statements} the best option seems to corresponds to the coarsest level of abstraction detail, while for \textit{node annotation}, \textit{global variables}, and \textit{global functions} the best option seems to corresponds to one of the intermediate levels of abstraction detail. For our future work, we aim to perform a deeper analysis on this and hopefully learn such configurations, to reduce (or eliminate) the overhead necessary of trying to manually discover such configurations.

%% file: related.tex
\section{Related Work}

Research into code representations in the space of code similarity is still in its infancy, yet, there is a large and growing body of work to consider. A classical approach to infer code semantics is to utilize an intermediate representation (IR), such as LLVM~\cite{lattner:2004:cgo}. However, these representations were originally designed with the purpose of mapping efficiently to low-level instruction set architectures (ISAs). As such, they might not be ideal candidates for code similarity. 

Still, advances in ML seem to have stimulated a number of approaches~\cite{ben-nun:2018:neurips, zhao:2018:esec/fse} that rely on such IRs to infer high-level semantics for the purpose of code similarity. Nevertheless, such approaches suffer from the disadvantage of requiring compilation to determine the validity of the input code, hence limiting their applicability. Other research avoids this reliance on compilation by representing a program using its dynamic execution trace~\cite{wang:2018:iclr}. While such approaches enable the encoding of concrete details of the program semantics, the collection of dynamic traces can be costly (program execution is required). 

The idea of utilizing the compiler's IR has been extended further by more recent ML-based approaches~\cite{alon:2019:popl, alon:2019:iclr, zhang:2019:icse}, that use an \textit{abstract syntax tree} (AST), which is at a higher level of abstraction than some IRs, such as LLVM. These AST approaches tend to rely on featurizing the AST to include some of its meta-properties, such as paths in the AST, to discover structural similarities. A key intuition for these approaches is that structural similarity of an AST may correlate to code similarity. Other recent research has focused on constructing code representations from raw source code tokens or sequences~\cite{sachdev:2018:mapl, sajnani:2016:icse} with some success. However, these rely on certain strict assumptions on the input code and might be challenging to generalize. 

To our knowledge, Aroma's simplified parse tree (SPT) represents a state-of-the-art structural representation for code similarity~\cite{luan:2019:oopsla}. Aroma extends all the aforementioned approaches in that it uses a customized parse tree representation, SPT, which encapsulates high-level code semantics. The SPT is at higher level of abstraction than previous AST-based approaches as it avoids representing irrelevant syntactic information. Our work is inspired by Aroma's SPT and aims to take it one step further by allowing for systematic exploration of a range of customizable configuration parameters that control CAPT's construction.

%% file: conclusion.tex
\vspace{-1mm}
\section{Future Work and Conclusion}

In this paper, we presented the context-aware parse tree (CAPT), a novel tree structure that we have developed principally for the purpose of code similarity analysis. CAPT is heavily inspired by Aroma's simplified parse tree (SPT). 

Our research quantitatively demonstrates the value of our proposed semantically-salient features, enabling a specific CAPT configuration to be 39\% more accurate than SPT across the 48,610 programs we analyzed. We believe CAPT is able to produce improved code similarity accuracy because it provides a more flexible semantic configuration across \emph{(i)} language-specific ambiguity resolution and \emph{(ii)} unbinding support via language-agnostic techniques for removal of syntactic features that are semantically irrelevant. Our exploration into CAPT is still in its infancy.